\documentclass[prb,twocolumn,showpacs]{revtex4}
\usepackage{graphicx,psfrag,amssymb,amsmath}
\begin{document}
\title{Local phonon mode in a fermionic bath, and its relation to the Kondo effect}
\author{Bal\'azs D\'ora}
\email{dora@mpipks-dresden.mpg.de}
\affiliation{Max-Planck-Institut f\"ur Physik Komplexer Systeme, N\"othnitzer Str. 38, 01187 Dresden, Germany}

\date{\today}

\begin{abstract}
We have studied the interplay of a local phonon mode embedded in a metallic host (Holstein impurity model) using 
Abelian bosonization. The phonon 
frequency softens, which takes place in two steps: first, their frequency starts softening, and acquires finite
lifetime. Then oscillations disappear from the response, and 
two distinct, finite dampings characterize them.  
Similar behaviour shows up in the spin-boson model.
Due to phonons, the electrons experience an attractive, dynamic interaction. As a result, 
the electronic charge response 
enhances similarly to the spin response in the Kondo model. The local electronic density of states develops a dip 
around zero frequency. 
Thus the chance of charge-Kondo effect emerges.

\end{abstract}


\maketitle

\section{Introduction}

Single impurity models have a long history\cite{hewson}, and our understanding has profited a lot from different 
reliable techniques. 
Numerical renormalization group, conformal field theory, Nozi\`eres' Fermi liquid description, Bethe ansatz solution and 
bosonization completed each other, and highlighted different aspects of the same problem. In many cases, the latter, 
being nonperturbative and analytic, was able to provide us with a transparent picture of the underlying 
physics\cite{nersesyan}. Its 
appeal is due the 
elegance of analytic solution, being able to treat
dynamic
quantities as well (correlation functions), and the simplicity of the picture emerging from it.

Beyond electron-electron interaction, which is a basic ingredient in Kondo type models, another source of 
correlation is represented by phonons\cite{haldane,hewsonmeyer,cornaglia,mozyrsky,lundin,zhu}. They 
play a 
prominent role in explaining the conventional s-wave BCS superconductivity and the charge density wave formation in low 
dimensional systems. Their general feature is the generated dynamic interaction between the electrons, which, in 
essence, is attractive for small  energy transfers. 

In many strongly correlated systems, such as heavy fermions or valence fluctuation systems, lattice 
vibrations are known to couple strongly to electrons. The recent discovery of magnetically robust heavy fermion 
behaviour in filled skutterudite compound\cite{sanada} SmOs$_4$Sb$_{12}$ renewed interest in Kondo phenomena with 
phononic 
origin\cite{yotsuhashi,hotta}. Both the specific heat coefficient and coefficient of the quadratic temperature 
dependence of the 
electrical resistivity
were
found to be almost independent of an applied magnetic field.
Quantum dots, especially the ones based on single molecules 
(C$_{60}$ for 
example\cite{yu}), also possess vibrational degrees of freedom, which will react to the electron transfer through 
them\cite{park}. 
Therefore the properties of impurity models dominated by phonons are challenging. The applied methods should be able 
to 
treat 
dynamical properties, which rules out techniques as the Bethe ansatz.

Several theoretical investigation focused on electron transport in the presence of electron-phonon interaction, mainly 
in the context of the single impurity Anderson-Holstein model. This model consists of an Anderson impurity with linear 
coupling to a local phonon mode\cite{hewsonmeyer}.  Hewson and Newns considered its spinless version with a few 
electron in the system without explicit 
electron correlation\cite{hewson1,hewson2}. Later further studies  
 based on 
the NRG (numerical renormalization group)\cite{hewsonmeyer,cornaglia,jeon,hotta} and the non-equilibrium Keldysh 
formalism\cite{lundin,zhu} have been performed.
These works reported about the softening of the local phonon mode, and the enhancement of the charge susceptibility. 
These phenomena point toward the realization of the charge-Kondo effect, 
caused by the degeneracy of zero and doubly occupied electron states. 

In the present work, using Abelian bosonization, for the first time to our knowledge to attack the local electron-phonon 
problem, we can not only confirm the prediction of previous works, 
but also study analytically the specific heat, the phonon Green's 
function, the charge susceptibility and especially the local density of states. 
We  
follow the softening of phonons from weak to strong coupling, and connect the present problem with the underscreened 
Kondo model in magnetic field.
As the electron-phonon coupling increases, phonon states with large occupation number start to play an important role 
even at very low temperatures, leading to a transition to a phonon distorted state only at infinitely strong coupling. 
Similar phenomenon reveals itself in the spin-boson model as well\cite{weiss,rmpleggett}.
Then, in contrast to NRG, where only a finite number of phonons can be taken 
into account, we can consider states with arbitrary phonons, and the crossover and the softening at arbitrary 
electron-phonon couplings.

\section{The model}

In real materials, electrons interact with each other explicitly through the Coulomb force and implicitly through  
phonons. 
In several cases, the resulting phase is well described by Landau's Fermi liquid theory with renormalized quasiparticle 
parameters due to interactions\cite{abrikosov}.
However, certain inhomogeneities like impurities, lattice imperfections are usually present, and exhibit a source of scattering. 
When these are of electronic origin, spin-Kondo physics is evoked\cite{hewson}.
The local perturbations can also show up in the form of local lattice vibrations. 
This is why we have undertaken the study of spinless fermions interacting with a single, dispersionless Einstein phonon 
mode, mainly at zero 
temperature. This can be called the single impurity Holstein model, which is related to the Holstein model as the 
single impurity Anderson model is 
related to its periodic version.
It also parallels to the one studied in Ref. \onlinecite{hewson1,hewson2} in the same way as the single 
impurity Wolff model\cite{wolff} is connected with  the Anderson model\cite{hewson}. 
A similar model has been studied\cite{yotsuhashi} describing the coupling of a local Einstein phonon to conduction 
electrons by the numerical renormalization group, with emphasis on the nature of the fixed points, and relations 
with the two-level Kondo effect have been revealed.
The present model can also be relevant for the dynamical mean-field theory (DMFT) of the Holstein lattice 
model\cite{capone}, which maps it onto the Holstein impurity model.
Coupling to acoustic 
phonons would also be a fruitful proposition, but the essence of physics is readily captured by the simplest model.
When the coupling of the electron density to the phonon displacement field is isotropic, the model can be mapped onto a 
single branch of chiral fermions interacting with a single Einstein phonon only at the origin, and is suitable for 
Abelian bosonization. This mapping follows closely the one proposed in Ref. \onlinecite{affleck2} for the Kondo 
model\cite{nersesyan}.

In the Hamiltonian language, the model is given by:
\begin{gather}
 H=-iv\int\limits_{-L/2}^{L/2} dx
\Psi^+(x)\partial_x\Psi(x)+gQ\rho(0)+\frac{P^2}{2m}+\frac {m\omega_0^2}{2}Q^2,
 \label{hamilton}
\end{gather}
and only the radial motion of the particles is accounted for by chiral (right moving) fermion
field\cite{affleck2}, $\rho(x)=:\Psi^+(x)\Psi(x):$ is the normal ordered electron density, $L$ is the length of the 
system, $v$ is the 
Fermi 
velocity, $g$ describes the local electron-phonon coupling, $m$ and $\omega_0$ are the phononic 
mass and frequency, respectively, $Q$ and $P$ are the phonon displacement field and momentum conjugate to 
it. 

In spirit, this model is similar to the 
underscreened 
Kondo model in a magnetic field (which quenches the remaining degrees of freedom) with phonons replacing the 
impurity 
spin, since the phonon displacement field 
(Q) can take any real values, while the electron density trying to compensate it, is bounded, although states with a large number of bosons in the harmonic oscillator hardly contribute to the physics at low temperatures due to their high energy. The underscreened Kondo 
model in a magnetic field is governed by a Fermi liquid fixed point. We speculate that the 
model under study 
produces similar behaviour. When the magnetic field is switched off, the underscreened Kondo
model shows singular Fermi liquid behaviour\cite{colemanpepin}, whose analogue in the present case occurs at 
$g\rightarrow \infty$.  
Since the charge degrees of freedom are coupled to the local bosons, the possibility of 
observing the "underscreened" charge-Kondo effect opens\cite{haldane}.
The realization of the charge-Kondo effect has also been found in similar models\cite{taraphder,zitko}.
Our model is similar to the one describing a tunneling particle coupled to a fermionic environment\cite{waxman}.
The model can be bosonized via\cite{shankar,delft}
\begin{equation}
\Psi(x)=\frac{1}{\sqrt{2\pi\alpha}}\exp{(i\sqrt{4\pi}\Phi(x))}
\label{fermionboson}
\end{equation}
to lead 
\begin{equation}
H=v\int\limits_{-L/2}^{L/2} dx
(\partial_x\Phi(x))^2+\frac{g}{\sqrt\pi}Q\partial_x\Phi(0)+\frac{P^2}{2m}+\frac 
12 m\omega_0^2Q^2.
\label{hamboson}
\end{equation}
If we integrated out the phonon degrees of freedom in Eq. \eqref{hamilton} or \eqref{hamboson}, 
we would arrive to a local interaction  
between 
electrons, given by ${2g^2}/{m(\omega^2-\omega_0^2)}$, $\omega$ is the energy transfer in the interaction. 
At low energies ($|\omega|<\omega_0$), this would lead to an effective attractive interaction, and the resulting 
Hamiltonian would 
coincide with the spin sector of the repulsive Wolff impurity 
model\cite{wolff,zhang,dorawolff}, 
responsible for the spin-Kondo phenomenon. Hence our model is capable to show Kondo physics in the charge sector. 
This further strengthens our proposal for the charge-Kondo effect.
Attractive local 
interactions has already provided us 
with similar phenomenon\cite{hewsonmeyer,cornaglia}.

This resulting Hamiltonian (Eq. \eqref{hamboson}) is similar to the Caldeira-Leggett Hamiltonian (CL), 
describing the dynamics of a particle coupled 
to dissipative environment\cite{caldeira}. However, the roles are reversed between oscillators and 
tunneling particle in Eq. \eqref{hamboson} and in Ref. \onlinecite{caldeira}: our localized phonon represents the 
tunneling particle in CL language, while the bosonized ($\Phi$) 
fermionic field stands for the dissipative environment of CL, represented by harmonic oscillators. 
An important difference between the two models is the lack of renormalization of potential in CL, ensuring, that the 
coupling to reservoir solely introduces dissipation\cite{weiss}.
In our case, however, we expect the softening of the phonon mode due the interaction with fermions on physical ground,
hence the renormalization of the phonon frequency plays an essential role as is testified later in Eq. 
\eqref{fononenergia}, in addition to dissipation. 
Moreover, the properties of our original fermionic field must be obtained through Eq. \eqref{fermionboson}, a 
difficulty which is never faced with in the Caldeira-Leggett model.

Our bosonized Hamiltonian (Eq. \eqref{hamboson}) also shares similar properties with the spin-boson model, where a 
two-level system is coupled to a bosonic environment\cite{rmpleggett,weiss}. There, assuming an Ohmic bath, at a finite 
value of the coupling of 
the boson-two-level system, a quantum phase transition has been identified. In our case, the bosonic bath is represented
by the bosonized fermions, but the two-level system is replaced by a multi-level system, i.e. a harmonic oscillator.
As a result, the obtained behaviour of our model differs from that found in the spin-boson model, although some 
dynamical features, connected with the oscillation frequency, are similar.   

\section{Limiting cases}
\label{limit}

The adiabatic limit ($\omega_0\rightarrow 0$) is meaningful, if it is taken together with $m\rightarrow\infty$, keeping 
$m\omega_0^2=\kappa$ finite.
The phonons kinetic energy vanishes, and only a static displacement is felt by the electrons. Hence, the model 
reduces to a simple scattering problem, although one has to minimize the total free energy of the system 
(electronic contribution  + elastic energy $\kappa Q^2/2$) with respect to the 
frozen phonon displacement field to obtain the correct solution, which transforms our problem into a 
self-consistent 
mean-field theory, which is the exact solution in this limit.
The free energy, subject to minimization with respect to $Q$, reads at $T=0$ as
\begin{equation}
\Delta F=-n_0gQ\frac 2\pi \int\limits_0^1 d\lambda\arctan(\lambda \pi gQ\rho)+\frac \kappa 2 Q^2,
\end{equation}
which leads to the self-consistency equation for $Q$:
\begin{gather}
\frac{\pi \kappa}{2gn_0}Q=\arctan(\pi gQ\rho),
\end{gather}
where $n_0=W\rho$ is the average electron density per site. In the weak coupling limit, only the trivial solution exists 
($Q=0$), as opposed to standard mean-field treatments of the Holstein lattice model\cite{hirsch}. As $g$ exceeds 
$g_c=\sqrt{\kappa/2n_0\rho}$, a non-trivial 
solution shows up, which minimizes the free energy. This behaves as $Q=\sqrt{6(g-g_c)}/\pi\rho g_c^{3/2}$ close to 
$g_c$, and $Q=gn_0/ \kappa$ as $g$ grows to infinity.
For $g>g_c$, the electrons feel a local scattering center with strength $V=gQ$, and the 
local electronic density of states 
is expected to be suppressed\cite{mahan} close to zero frequency as $1/(1+(\pi V\rho)^2)$. Complete suppression occurs 
only for $g\rightarrow \infty$.
Phonons lose their dynamics 
completely.
However, when quitting the strict adiabatic limit, the displacement field is not frozen any more, 
quantum corrections are expected to destroy ordering. 
In essence, this is similar to the large N limit of impurity models\cite{hewson} (N denotes the spin degeneracy), where a true phase transition occurs only in the 1/N=0 limit.
Note, that similar calculation applies to the adiabatic limit of the Anderson-Holstein impurity model as well.

In the anti-adiabatic limit, $\omega_0\rightarrow\infty$ while keeping the 
ratio $g/\omega_0$ finite, the phonons react instantaneously to the electrons, leading to non-retarded interaction 
between 
the densities as
\begin{equation}
H_{int}=-\frac{g^2}{2m\omega_0^2}(n-n_0)^2,
\end{equation}
where $n$ is the charge density of electrons at the origin. Had 
we chosen spinful fermions, the 
resulting model would coincide with the negative $U$ Wolff model\cite{wolff,dorawolff} ($U=-g^2/m\omega_0^2$). In the 
spinless case, this 
again reduces to potential scattering with strength $V=g^2(2n_0-1)/2m\omega_0^2$. Again, the local 
density 
of states is expected to be suppressed close to zero frequency, similarly to the adiabatic case.

Finally by taking the atomic or zero bandwidth limit, which also describes the infinite coupling case 
($g\rightarrow\infty$), the phonons couple only to an isolated electron:
\begin{gather}
H_{at}=c^+c (E+gQ)+\frac{P^2}{2m}+\frac{m\omega_0^2}{2}Q^2,
\end{gather}
where $E$ is the c-level energy. This model is known as the independent boson model\cite{mahan}. Since the electron 
number operator ($c^+c$) is conserved, most of the physical quantities can trivially be evaluated. However, 
the electron Green's function does not belong to this class. After a unitary transformation 
($U=\exp(-igPc^+c/m\omega_0^2)$), this can be evaluated exactly as
\begin{gather}
G_{at}(t)=\langle c(t)c^+\rangle=(1-\langle 
c^+c\rangle)\times\nonumber\\
\times\exp\left[-i\left(E-\frac{g^2}{2m\omega_0^2}\right)t+\frac{g^2}{m\omega_0^2}\left(D_Q(t)-\langle 
Q^2\rangle\right)\right].
\label{atomic}
\end{gather}
Here the first term is the renormalized c-level Green's function, $\langle
Q^2\rangle=1/2m\omega_0$ is the mean square value of the displacement field, being independent of the coupling in this 
case. Finally, $D_Q(t)$ is the Green's function of the displacement field $Q$ given by 
\begin{equation}
D_Q(t)=\langle Q^2\rangle \exp(-i\omega_0 t).
\end{equation}
From this, one can evaluate the density of states, which reads as
\begin{equation}
\rho_{at}(\omega)=\exp(-\gamma)\sum_{n=0}^\infty\frac{\gamma^n}{n!}\delta\left(\omega-E+\gamma\omega_0-n\omega_0\right)
\end{equation}
with $\gamma=g^2/2m\omega_0^3$. 
Compared to the original density of states with $g=0$, where a single Dirac-delta peak contains all the spectral weight,
it is now distributed non-uniformly among infinite number of peaks. 
In general, most of the spectral weight is concentrated around $\omega\approx E$ for a finite $g$ as well.
In the weak coupling limit ($\gamma\ll 1$), the weight decreases monotonically with $n$.
For strong coupling ($\gamma\gg 1$), the largest weights are still distributed around 
$\omega\approx E$, which means that the $n\approx\gamma$ terms are the most dominant in the sum. These weights are 
proportional 
to $1/\sqrt{n}$. From this, one can conclude, that by increasing $g$ to move from weak to strong coupling, 
terms with $n\propto \gamma$ contain the largest spectral weights, which, however decreases as $1/\sqrt{\gamma}$. This 
means that
increasing amount of spectral weight is transfered away from the central region, hence the local density of states is 
suppressed due to phonons, as borns out from the above three limiting cases. It is only suppressed to zero for infinitely 
strong coupling $g$.

\section{Softening of the phonon}

To determine the dynamics of our system, we start with the evaluation of the Green's function of the 
phonon displacement field, defined by 
\begin{equation}
D_Q(\tau)=\langle T_\tau Q(\tau)Q(0)\rangle,
\end{equation}
whose Matsubara form can easily be calculated using the standard diagrammatic technique from Eq. \eqref{hamboson}, which 
yields to
\begin{equation}
D_Q(i\omega_m)=\frac 1m \frac{1}{\omega_m^2+\omega_0^2-2\rho g^2\chi(\omega_m)/m},
\label{fonongreen}
\end{equation}
where $\rho=1/2\pi v$, $\omega_m$ is the bosonic Matsubara frequency and 
\begin{equation}
\chi(\omega_m)=\sum_{q>0}\frac{(vq)^2}{\omega_m^2+(vq)^2}
\end{equation}
is the local susceptibility of the fermions without the phonons. After analytic continuation to real frequencies, it is 
given by $\chi(\omega\ll W)\approx 
\rho (W+i\pi\omega/2)$ for small frequencies, $W$ is the cutoff or bandwidth, and the obtained formula holds 
regardless to the 
chosen cutoff procedure.
Our Eq. \eqref{fonongreen} is exact for the Hamiltonian given by Eq. \eqref{hamboson}, no other diagrams are left to 
consider. However, the bosonization cutoff scheme renormalizes the coupling constant\cite{nersesyan}, and the impurity 
potential $g$ needs to be replaced by the appropriate scattering phase shift. Since this is not a universal 
quantity (apart from the linear term), it depends on the regularization scheme, and we consider it to be replaced by the 
renormalized one onwards, 
as is standard in impurity models\cite{nersesyan}.
 The dynamics of the phonons can be inferred by investigating the pole structure of Eq. \eqref{fonongreen}.
The excitations energies become complex (i.e. finite lifetime or damping of phonon excitations) in the presence of 
finite coupling to the electrons, and are given 
by
\begin{equation}
\omega_{p\pm}=-i\Gamma\pm\sqrt{\omega_0^2-\Gamma^2-4\Gamma W/\pi} 
\label{fononenergia}
\end{equation}
with $\Gamma=\pi(g\rho)^2/2m$. Below $\Gamma<\Gamma_1=-2W/\pi+\sqrt{4W^2/\pi^2+\omega_0^2}$, the 
square root in Eq. \eqref{fononenergia} is real, hence the  
displacements reach their equilibrium in an oscillatory fashion within a time characterized by 
$-1/$Im$\omega_p$.
For higher frequencies, the phonon mode is completely softened, Re$\omega_p=0$ (the square root becomes 
imaginary), the excitations have two different finite lifetimes or dampings determined from Eq. 
\eqref{fononenergia} between $\Gamma_1<\Gamma<\Gamma_2=\pi\omega_0^2/4W$. 
Interestingly, the very same critical value $\Gamma_2$ would be found, if we inserted the value of 
$g_c=\omega_0\sqrt{m/2n_0\rho}$ found in 
the extreme adiabatic limit (or self-consistent mean field theory) to $\Gamma=\pi(g\rho)^2/2m$, which is, in turn, zero 
because $\omega_0\rightarrow 0$.
In this range, phonons can be excited with zero energy, and all 
the displacements are relaxed to equilibrium without oscillations. This relaxation slows down  
close to $\Gamma_2$. 
This region is very narrow, because for realistic values, $\omega_0\ll W$, it shrinks as
$\sim\omega_0^4/W^3$. Similar relaxation characterizes two-level systems coupled to a bosonic 
bath\cite{rmpleggett,weiss}.
For higher values of $\Gamma$, our 
approach breaks down, it signals lattice distortion with $\langle Q\rangle \neq 0$, as is indicated by 
the complete softening of the phonon mode. However, such a 
phenomenon is impossible in zero dimensional systems, and we ascribe it to bosonization. 
The explicit functional form of physical quantities on model parameters obtained by bosonization can 
deviate from the exact 
one. The phase shifts in impurity problems\cite{zaranddelft}, the correlation exponents in Luttinger liquids determined 
via 
bosonization are only correct in the weak coupling limit\cite{nersesyan}.
This and the previous argumentation suggests, that in reality, $\Gamma_2$ should only be reached for $g\rightarrow\infty$, 
but the complete softening predicted at $\Gamma_1$ would take place at a finite value of $g$, and $\Gamma\sim g^2$ only 
at small $g$.
In the adiabatic limit, ($m\rightarrow\infty$, $\omega_0\rightarrow 0$) $\Gamma$ also vanishes, and $\omega_{\pm}=0$.

We mention that such softening remains absent in the Caldeira-Leggett model due to the compensation of the 
potential renormalization there\cite{caldeira}.
Similar phenomenon occurs in a two-level system, coupled to a bath of three dimensional phonons\cite{silbey}.
The renormalized tunneling rate drops to zero only at infinitely strong coupling.

The softening of the phonon frequency was also found in the NRG treatment of the Anderson-Holstein impurity 
model\cite{hewsonmeyer,cornaglia,jeon}, but this has only been followed in a moderate range of parameters. 
DMFT studies of the Holstein lattice model also revealed similar features\cite{capone}.

\begin{figure}[h!]
\centering{\includegraphics[width=7cm,height=7cm]{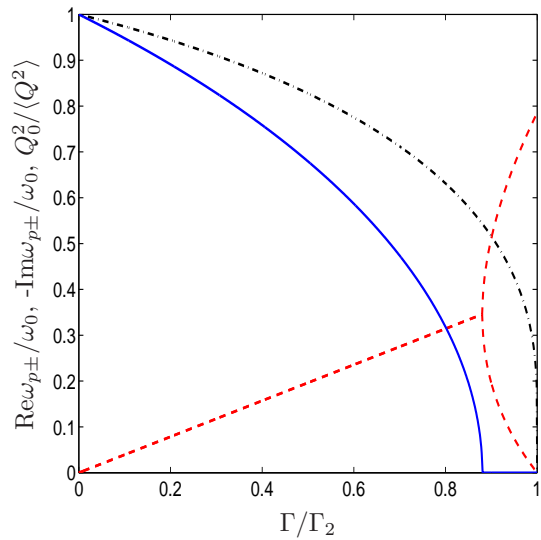}}
\caption{(Color online) The real (blue solid line) and imaginary (red dashed line) part of the phonon 
excitation energies, and the inverse of the mean square value of the displacement field (black dashed-dotted line) are 
visualized as a function of 
the 
electron-phonon coupling for $W/\omega_0=2$. 
At $\Gamma_1$, the phonon modes is softened, but the 
excitations have finite lifetime. At $\Gamma_2$, the damping vanishes, the lifetime 
($\sim -1/$Im$\omega_p$) diverges, 
and so does 
the mean square value of the 
displacement field. 
\label{2freq}}
\end{figure}

The general behaviour of the eigenfrequencies is sketched in Fig. \ref{2freq}. 
The phonon density of states contains two Lorentzians with resonance widths -Im$\omega_{p\pm}$, centered around  
Re$\omega_{p\pm}$. With increasing $\Gamma$, 
they become centered around the origin, and close to $\Gamma_2$, the one determined by $\omega_{p+}$ 
tends to $\delta(\omega)$. 
This is reminiscent to what happens in the channel anisotropic two channel Kondo model\cite{fabrizio}.
In comparison with the underscreened Kondo model in a magnetic field, a similar divergence shows up in the localized 
electron density of states\cite{colemanpepin} with the offset of magnetic field. The very same role is played in our 
case by $\Gamma\rightarrow\Gamma_2$, which causes the divergence as $\omega_{p+}\rightarrow 0$.
Softening of the phonon mode has been found by the numerical renormalization group as well\cite{hewsonmeyer} in a 
similar model. 

\section{Thermodynamics}

The mean value of the displacement field is zero. The mean square value of the displacement field is 
calculated from Eq. \eqref{fonongreen} at $T=0$ as
\begin{equation}
\langle Q^2 \rangle 
=\lim\limits_{T\rightarrow 0}T\sum_m 
D_Q(\omega_m)=Q_0^2\frac{2i\omega_0}{\pi(\omega_{p+}-\omega_{p-})}\ln\frac{\omega_{p-}}{\omega_{p+}},
\end{equation}
where $Q_0^2=1/2m\omega_0$ is the value for $g=0$. It is plotted in Fig. \ref{2freq}. As $\Gamma$ approaches $\Gamma_2$, 
it diverges as
\begin{equation}
\frac{\langle Q^2 \rangle}{Q_0^2}\approx 
\frac{\omega_0}{\pi\Gamma_2}\ln\left(\frac{4\Gamma_2^3}{\omega_0^2(\Gamma_2-\Gamma)}\right),
\end{equation}
indicating the change in the $\langle Q\rangle=0$ relation. 
This is traced back to the transition to the phonon distorted state at $g=\infty$, as discussed below Eq. 
\eqref{fononenergia}. Similar phenomenon has been observed in the spin-boson model\cite{rmpleggett,weiss} at a finite 
value of the spin-boson coupling, assuming an Ohmic bath. However, for a super-Ohmic bath, the transition occurs only 
at infinitely strong coupling\cite{silbey}.
In contrast, in Caldeira-Leggett-type Hamiltonians, dissipation induced squeezing occurs\cite{ambegaokar}, which 
originates from the 
effect of damping in reducing the rate of escape from a metastable well.

In a linear harmonic oscillator, the ground state wave function is a gaussian. It is natural to ask, to 
what extent this picture holds in the case of finite electron-phonon coupling. The probability 
distribution of the oscillator coordinator or displacement reads as
\begin{equation}
|\Psi_{osc}(x)|^2=\langle \delta(Q-x)\rangle=\lim_{a\rightarrow 
0} \int\limits_{-\infty}^\infty \frac{dk}{2\pi}\langle \exp(ik(Q-x)-ak)\rangle.
\end{equation}
In the second step, a useful representation of the Dirac-delta function was inserted.
Since the Hamiltonian is quadratic, or the corresponding action is gaussian for the phonons after 
integrating out the fermions, the expectation value of the exponent can be calculated following Ref. 
\onlinecite{delft}, which leads to
\begin{equation}
|\Psi_{osc}(x)|^2=\frac{1}{\sqrt{2\pi\langle Q^2\rangle}}\exp\left(-\frac{x^2}{2\langle 
Q^2\rangle}\right).
\end{equation} 
The ground state wave function remains gaussian, but the variance ($\langle Q^2\rangle$) increases 
monotonically with $g$. The region where the phonons are mainly restricted to, is wider than without 
electron-phonon coupling. 
The coupling to electron increases the average oscillator displacement around the equilibrium. 
This further corroborates the picture emerging from the previous studies. 

\begin{figure}[h!]
\centering{\includegraphics[width=7cm,height=7cm]{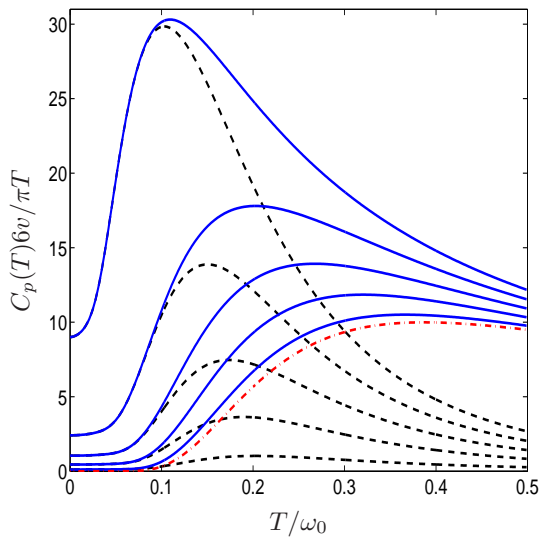}}
\caption{(Color online) The total (blue solid line) phonon specific heat contribution for different 
values of the coupling 
constant $\Gamma/\Gamma_2=0.1$, 0.3, 0.5, 0.7 and 0.9, $W/\omega_0=10$ from bottom to top. The red dashed 
dotted line shows the pure phononic $\Gamma=0$ specific heat, while the black dashed curves stands for their 
difference due to the finite coupling $g$. Note the sharpening of the broad bump at low $T$ as $\Gamma$ increases.
\label{phspec}}
\end{figure}

The phonon contribution to the free energy 
can be calculated by using Pauli's trick of integrating over the coupling constant. After some algebra, one 
arrives at
\begin{gather}
\Omega=\Omega_e+T\ln\left[2\sinh\left(\frac{\omega_0}{2T} \sqrt{1-\frac{\Gamma}{\Gamma_2}}\right)\right]+\nonumber\\
+
\int\frac{dx}{2\pi}b(x)\tan^{-1}\left(\frac{2\Gamma 
x}{x^2-\omega_0^2(1-(\Gamma/\Gamma_2))}\right),
\end{gather}
where $\Omega_e$ is the free energy of the fermions in the absence of phonons, $b(x)$ 
is the 
Bose distribution function, and the integral should be limited to the bandwidth, but this can be sent to 
infinity when obtaining quantities by differentiating $\Omega$\cite{fabrizio}. One has to carefully chose the appropriate 
phase angle of the $\tan^{-1}$ function. 
This expression is similar to the free energy of the two channel Kondo model in a magnetic field ($\sim 
\omega_0\sqrt{1-(\Gamma/\Gamma_2)}$ here) along the 
Emery-Kivelson 
line\cite{emery}, 
after replacing the Bose distribution function with the Fermi one. However, our "magnetic field" cannot be switched off, 
 hence the entropy is zero at $T=0$, similarly to the underscreened Kondo model\cite{sacramento}.
The specific heat can be calculated by $C(T)=-1/T(\partial^2\Omega/\partial T^2)$. Its low temperature ($T\ll
\omega_0\sqrt{1-(\Gamma/\Gamma_2)}$)
behaviour due to phonons reads as
\begin{equation}
C_{p}(T)=\frac{\pi T}{6v}\frac{\Gamma}{\Gamma_2-\Gamma},
\end{equation}
which sharpens when $\Gamma_2$ is approached.
The exponential freezing-out of the Einstein phonon changes to a linear $T$ dependence, and adds to  the 
Sommerfeld coefficient of the conduction electrons, and enhances it. Had we chosen acoustic phonons, their $T^3$ 
specific heat would also be overwhelmed at low temperatures. At higher  temperatures, a broad bump  
shows up in the specific heat as a function of temperature due to the presence of $g$.
 This can serve 
as an identifier of the local electron-phonon interaction, and is shown in Fig. \ref{phspec}.
Similar structures have been observed in the Anderson-Holstein impurity model by NRG\cite{hotta}.

\section{Electronic properties: susceptibility and local density of states}

The phonons are expected to have a profound impact on the electronic properties. Had we integrated them out, we 
would have arrived to a local dynamic electron-electron interaction, attractive at low energies and repulsive 
for higher ones. Static interactions already modify significantly the electronic response, as was demonstrated 
in the Wolff model\cite{wolff,schlottmann,dorawolff}.
In the followings we are going to see the effect of dynamic one explicitly on the electron system.

The electric charge is defined through $\rho(x)=\partial_x\Phi(x)/\sqrt\pi$. The local charge correlation function is evaluated from 
Eq. \eqref{hamboson} using the diagrammatic technique, and along the real frequency axis it reads as
\begin{equation}
\chi_{charge}(\omega)=\dfrac{2\rho\chi(\omega)}{1+\dfrac{2g^2\rho}{m(\omega^2-\omega_0^2)}\chi(\omega)}.
\label{chicharge}
\end{equation}
The local charge excitation energies are determined from the poles Eq. \eqref{chicharge}, which possess finite lifetime, as given by 
Eq. \eqref{fononenergia}. 
Eq. \eqref{chicharge} is the standard result in the random-phase-approximation, which turns out to be the exact one in the present 
case. Most of the many 
body contributions are canceled by the Ward identity, which relates the vertex function and the electron propagator. 
This could 
be expected from the fact that the model is solvable by bosonization. The very same phenomenon revealed itself during 
the study of the Wolff impurity model\cite{zhang,schlottmann}.
The dynamic nature of the phonons is observable in the denominator of Eq. \eqref{chicharge}. In the static limit, the 
effective interaction between the electrons is attractive, and changes to repulsive for frequencies exceeding the 
phonon energy 
$\omega_0$.
The static limit of the charge susceptibility simplifies to
\begin{equation}
\chi_{charge}(0)=\frac{2\rho^2W}{(1-(\Gamma/\Gamma_2))}.
\end{equation}
In accordance with Eq. \eqref{fononenergia}, this also predicts the transition to the distorted phase, 
which would be accompanied by the rearrangement of the charges as well.
This signals that our model is on the brink of charge Kondo effect.
Similar enhancement of the charge response has been reported in Ref. \onlinecite{hotta}.

Since we work with spinless electrons, the single particle Green's function can be evaluated, unlike in the SU(N) 
Wolff model\cite{wolff,dorawolff} or 
the two channel Kondo model\cite{emery}. 
There, such a calculation would involve formally $\sqrt{\Psi(x)}$, which is difficult to work with.
The local retarded Green's function is defined as
\begin{gather}
G_R(t)=-i\Theta(t)\langle\{\Psi(t),\Psi^+(0)\}\rangle=\nonumber\\
=-i\Theta(t)\frac{\exp(-4\pi\langle 
\Phi^2\rangle)}{2\pi\alpha}\left[\exp(4\pi 
C(t))+\exp(4\pi C(-t))\right],
\label{green1}
\end{gather}
where in the second step we made use of the gaussian nature of the action and used the usual tricks of operator 
manipulation\cite{delft,shankar}. The task is to evaluate the correlator 
$C(t)=\langle\Phi(t)\Phi(0)\rangle$, whose second appearance in Eq. \eqref{green1} follows from time reversal symmetry. 
Given the fact that the bosonized Hamiltonian (Eq. \eqref{hamboson}) is quadratic, we can evaluate 
this expectation value at 
bosonic Matsubara frequencies using the path integral representation of the problem\cite{dorawolff}. Then by 
transforming it to imaginary times, finally we can read off $C(t)$ after careful analytic 
continuation to real times.
First we find for the Matsubara form that
\begin{equation}
C(\omega_m)=\frac{1}{4|\omega_m|}+\frac{m}{2}\Gamma D_Q(\omega_m).
\label{cgreen}
\end{equation}
Here the first term is responsible for the $1/\tau$ decay of fermionic correlations. Fortunately, the second 
expression, accounting 
for the phonon contribution, is separated from the first one, and does not require ultraviolet regularization unlike the first one, 
where the large $\omega_m$ part need to be cut off to avoid divergences.

Compared to the exact expression derived in the atomic limit ($G_{at}(t)$ in Eq. \eqref{atomic}), Eqs. 
\eqref{green1}-\eqref{cgreen} have identical structures. It contains the free fermionic contribution, and in addition, 
the exponentiated phonon Green's function. Since the coupling constant appears in the exponent, together with $D_Q$, 
similarly to the exact 
solution in the atomic limit, this suggests the 
non-perturbative nature of our bosonization approach in the presence of a finite electron band. Hence we believe, that 
our study captures correctly the physics in the whole parameter range.

\begin{figure}[h!]
\vspace*{3mm}
\includegraphics[width=7cm,height=7cm]{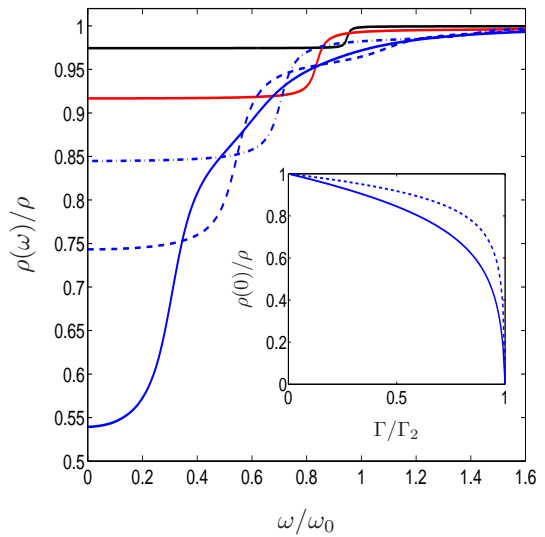}


\caption{The change of the local electron density of states at the impurity site is shown for $W/\omega_0=10$ from 
weak to strong coupling for 
$\Gamma/\Gamma_2$=0.1 (black), 0.3 (red), 0.5 (blue dashed-dotted), 0.7 (blue dashed) and 0.9 (blue solid) from top to 
bottom.
Increasing amount of spectral weight is transfered away from the zero frequency region with $\Gamma$.
  The inset visualizes the change in the residual 
density of states for $W/\omega_0=10$ (solid) and 20 (dashed), which vanishes at $\Gamma=\Gamma_2$.}
\label{dosomega}
\end{figure}

Its Fourier transform with respect to $\omega_m$ yields to the imaginary time ordered expression for the correlator, 
from which the 
desired relation can be obtained as
\begin{equation}
C(t)-C(0)=\frac{1}{4\pi}\ln\left(\frac{\alpha}{\alpha+ivt}\right)+\frac m2 \Gamma (D_Q(t)-\langle Q^2\rangle),
\end{equation}
where the equal time correlator need to be subtracted to regularize the first term on the right hand side and 
\begin{gather}
D_Q(t)=\frac{2i}{m\pi(\omega_{p+}-\omega_{p-})}\left(f(\omega_{p+}t)-f(\omega_{p-}t)\right)
\end{gather}
with
\begin{equation}
f(x)=-\sin(x)\textmd{Si}(x)-\frac \pi 2\sin(x)-\cos(x)\textmd{Ci}(-x),
\end{equation}
where Si$(x)$ and Ci$(x)$ are the sine and cosine integrals. 
At large times, its decay is characterized by the damping, Im$\omega_{p\pm}$, and the frequency of oscillations by 
Re$\omega_{p\pm}$.
By plugging this formula to $G_R(t)$, we evaluate the retarded single particle Green's function. The local density of states 
follows as
\begin{gather}
\rho(\omega)=-\frac 1\pi \textmd{Im}\int\limits_{-\infty}^\infty dt \exp(i\omega t)G_R(t),
=\nonumber \\ \rho\textmd{Re}\left(1+i\exp(-2\pi m\Gamma \langle Q^2\rangle )
  \int\limits_0^\infty \frac{dt}{\pi} \frac{\exp(i\omega t)}{t} \times \right.\nonumber \\
\times\left.\left[\exp(2\pi m\Gamma D_Q(-t))-\exp(2\pi m\Gamma D_Q(t))\right]\right).
\end{gather}
It is 
shown in Fig. \ref{dosomega} for various values of $\Gamma$.
As $\Gamma$ increases from weak ($\Gamma/\Gamma_2\ll 1$) to strong coupling ($\Gamma/\Gamma_2\leq 1$), the residual 
density of states decreases monotonically. In addition, increasing 
amount of spectral weight is transfered away from the central region around $\omega=0$, in perfect agreement with what 
we found in limiting cases of the model in Sec. \ref{limit}. Complete suppression at $\omega=0$ occurs in all cases at 
infinitely strong coupling, similarly to the bosonized case at $\Gamma\rightarrow\Gamma_2$. This is why we believe that
our approach captures correctly the strong coupling limit as well.
Close to the critical coupling $\Gamma_2$, the density of states exhibits a V-shapes form around the Fermi energy.
Similar redistribution of spectral weight was identified in the adiabatic, anti-adiabatic and atomic limit.
These findings are also in accordance with those found by NRG in the Anderson-Holstein impurity 
model\cite{hewsonmeyer,cornaglia,jeon}. 
There, in the non-interacting version of the model, a resonance peak shows up at the f-level energy, which narrows 
significantly in the presence of phonons, and increasing amount of spectral weight is transfered away from this region.
A more direct comparison is difficult, since the  Anderson-Holstein impurity
model distinguishes between conduction and localized electrons, which are weakly 
hybridized with the conduction band, as opposed to our 
one-band model. Similar reasoning separates the purely electronic Anderson impurity model from the Wolff impurity
model\cite{schlottmann}.

The significant decrease (dip like structure)  in $\rho(0)$ should be detected in point contact spectroscopy or by 
scanning tunneling microscopy
experiments in bulk materials. 
For small values of $\Gamma$, a 
steplike drop occurs with decreasing frequency close to  $\omega=\omega_0\sqrt{1-\Gamma/\Gamma_2}$, which smoothens as 
the electron-phonon 
coupling increases, through the appearance of other steplike features, as is seen in Fig. \ref{dosomega} for 
$\Gamma=0.7 \Gamma_2$. 
As $\omega_0$ increases, the drop in the density of states occurs at higher frequencies ($\propto  
\omega_0\sqrt{1-\Gamma/\Gamma_2}$), and the suppression of the residual density of states at $\omega=0$ is more 
efficient for higher $\omega_0$, as can be seen in the inset of Fig. \ref{dosomega}.

These 
features are different from that induced by a non-magnetic impurity or 
by a magnetic one, and the reason is the effective, dynamically attractive interaction generated by the phonons.

\section{Conclusion}

In conclusion, we have studied a model of spinless fermions interacting with an Einstein phonon (single impurity 
Holstein model) using Abelian 
bosonization. With increasing coupling, the phonon mode softens, and at $g\rightarrow\infty$, distortion occurs.
The 
softening of the phonon resonance takes place in two steps: first, the frequency starts softening, and acquires finite 
lifetime. Then oscillations disappear from the response, and two distinct, finite dampings characterize the phonons. 
Finally, at a critical coupling, which is conjectured to be at $g=\infty$, the phonons are distorted.
These manifest themselves in the specific heat through the enhancement of the Sommerfeld coefficient. 
The phonon's entropy is similar to that of the Kondo model in finite magnetic field\cite{sacramento}, hence vanishes at 
$T=0$.
The present model resembles closely to the underscreened Kondo model in magnetic field. 
The local charge susceptibility is strongly enhanced, similarly to the spin susceptibility of the Kondo model. This 
points toward the realization of charge-Kondo effect.
Closed expression is derived for the local 
electronic density of states. With increasing coupling, significant amount of spectral weight is transfered away from 
the $\omega=0$ region, in accordance with results in the atomic limit and in the Anderson-Holstein impurity model.
This should be observable in point contact spectroscopy, or on a molecule trapped near a tunnel junction, and influence 
the current-voltage characteristic as well.

\begin{acknowledgments}

We thank P. Fazekas and P. S. Cornaglia for useful discussions.
This work was supported by the Hungarian
Scientific Research Fund under grant number OTKA TS049881.
\end{acknowledgments}

\bibliographystyle{apsrev}
\bibliography{wboson}
\end{document}